# Detecting triplet states in opto-electronic and photovoltaic materials and devices by transient optically detected magnetic resonance

Jeannine Grüne, Vladimir Dyakonov and Andreas Sperlich*

Experimental Physics 6, Julius Maximilian University of Würzburg,
Am Hubland, 97074 Würzburg, Germany

**Abstract**

Triplet excited states in organic semiconductor materials and devices are notoriously difficult to detect and study with established spectroscopic methods. Yet, they are a crucial intermediate step in next-generation organic light emitting diodes (OLED) that employ thermally activated delayed fluorescence (TADF) to upconvert non-emissive triplets to emissive singlet states. In organic photovoltaic (OPV) devices, however, triplets are an efficiency-limiting exciton loss channel and are also involved in device degradation. Here, we introduce an innovative spin-sensitive method to study triplet states in both, optically excited organic semiconductor films, as well as in electrically driven devices. The method of transient optically detected magnetic resonance (trODMR) can be applied to all light-emitting materials whose luminescence depends on paramagnetic spin states. It is thus an ideal spectroscopic tool to distinguish different states involved and determine their corresponding time scales. We unravel the role of intermediate excited spin states in opto-electronic and photovoltaic materials and devices and reveal fundamental differences in electrically and optically induced triplet states.

**New concepts**

Optically detected magnetic resonance (ODMR) is a widely used technique to probe paramagnetic spin states associated with luminescence. On the one hand, ODMR can be implemented as a continuous wave (cw) method based on lock-in detection, which, however, makes it difficult to determine the correct sign of the effect, signal amplitude and in general to separate superimposed signals. On the other hand, pulsed techniques such as pulsed EPR or pulsed ODMR are used to study coherent properties of optically accessible spin centers. In this paper, we present an intriguing intermediate case for the study of luminescent materials, where coherent spin manipulation is not readily applicable due to a rapid spin decoherence. We show that by applying "long" microwave pulses over milliseconds, ODMR intensities reach equilibrium conditions, while signal formation and decay can be observed with superior signal-to-noise ratio. The recording of time-dependent spectra enables the deconvolution of temporally superimposed contributions with correct amplitudes, signal signs and time constants. We demonstrate the setup, principle and analysis of this method on optically- and electrically-generated excited triplet states and present the advantages within two case studies of molecular donor:acceptor systems for OPV and OLED applications.

**I. Introduction**

The most common form of electron paramagnetic resonance (EPR) spectroscopy is detecting the interaction of microwave irradiation with electron spins in a magnetic field.[1] There are several different implementations of EPR, such as continuous wave (cw) EPR[2], pulsed EPR[3], transient (tr) EPR[4,5], double electron-electron resonance (DEER)[6] equivalent to pulsed electron electron double resonance (PELDOR)[7] and a number of others. When applied to luminescent materials or opto-electronic devices, the primary interest is on the paramagnetic species that directly affect the optical properties. Therefore, the focus is to probe only those spin states that are directly related to luminescence without measuring the background of unrelated paramagnetic radicals or charge traps.[1] For this purpose, optically detected magnetic resonance (ODMR) is a well-known and widely used technique that combines spin-sensitive and optical spectroscopy that has been performed since 1950.[8]

Apart from the crucial advantage that ODMR only examines the paramagnetic species directly associated with luminescence or absorption, optical detection allows to study short-lived photoexcitations not detectable in cwEPR and with higher sensitivity than transient EPR.[9] The spin states to be investigated with ODMR can be related to the analyzed luminescence in various ways. Either they are directly causing the luminescence or the paramagnetic spin states are indirectly correlated. The former is exemplified by phosphorescent materials[10-14] or photoluminescence of defect centers, e.g., NV centers in diamond[15]. The latter is represented by (reverse) intersystem crossing of non-emissive triplet states to fluorescent singlet states, which is often the case for triplet systems in organic semiconductors, such as in organic photovoltaic (OPV)[16-18] or organic light emitting diodes (OLEDs)[19, 20]. Applying resonant microwave irradiation disturbs the steady-state equilibrium of the triplet sublevels. This change in spin polarization leads to a change in luminescence, which is detectable with high sensitivity via ODMR.

Pulsed techniques, such as pulsed EPR or pulsed ODMR[21, 22], are used to determine spin relaxation times or study coherent spin effects by applying short, high-power microwave pulses.[23] However, these techniques require a certain coherence time to build up a detectable alignment of the involved spins. Therefore, organic systems in particular with short $T_2$ coherence times in the range of up to or less than hundreds of nanoseconds[24-27] often prevent the retention of the required spin coherence beyond the deadtime of the spectrometer and are thus difficult to study.[23]

We present a transient ODMR technique, using a comparable setup as for pulsed ODMR, that employs "long" microwave pulses up to milliseconds, as already performed in earlier works.[28-31] The aim is not to build up coherent spin alignment, but to manipulate spin polarisation between the triplet sublevels. Therefore, we can observe processes that occur on longer timescales and boost signal intensities with the long pulses to achieve superior signal-to-noise ratio. By sweeping the magnetic field, time-dependent ODMR spectra are recorded which yields reliable absolute signal intensities and sign in contrast to lock-in detection. Applying a two-dimensional global fit enables the separation and identification of superimposed spin-dependent contributions to luminescence or possible exciton loss processes of the system. Thereby, we demonstrate this method not only on optically induced spin states, but also on electrically excited states. In this way, we can reveal a detailed insight into the excitation pathways, which differ substantially for optically excited thin films and electrically driven devices.

In this work, transient ODMR is performed on materials emitting photoluminescence (transient photoluminescence (PL) detected magnetic resonance, trPLDMR) or electrically driven devices producing electroluminescence (transient electroluminescence (EL) detected magnetic resonance, trELDMR), such as OLEDs or reverse operated solar cells. We choose continuous optical or electrical excitation since the envisioned opto-electronic device applications are also based on continuous operation. In pulsed optical or electrical excitation mode, the densities of charge carriers or triplet excitons, as well as their interaction rates would differ decisevely, limiting the usefulness of pulsed techniques. We first introduce the setup and pulse sequence used for this technique to record two-dimensional PLDMR and ELDMR spectra. Subsequently, we provide two case studies of OPV and OLED materials as well as opto-electronic devices to demonstrate the advantages and usefulness of this technique.

## II. Methods

**Figure 1** depicts the setup and the following components that constitute a modified pulsed EPR spectrometer whereby luminescence transients are recorded instead of microwave echoes:

   i. A photodetector (Hamamatsu Si photodiode S2387-66R) behind a long-pass filter that blocks excitation light.

   ii. A current/voltage amplifier with bandwidth suitable for the time resolution of the system (Femto DHPCA-100) amplifies the photocurrent.

   iii. A 16-bit digitizer card records the photocurrent transients in DC coupling mode (GaGe Razor Express 1642 CompuScope).

   iv. A TTL pulse sequence generator (SpinCore PulseBlasterESR-PRO) triggers digitizer and microwave generator.

   v. A microwave generator (Anritsu MG3694C) produces pulses of a set length, amplified up to 44 W (53 dB gain) with a subsequent solid-state or traveling wave tube amplifier (TWTA: Varian VZX 6981 K1ACDK).

   vi. Electromagnet for external magnetic field $B_0$.

At this point, trPLDMR and trELDMR setups differ. trPLDMR (**Figure 1a**) uses thin film samples in EPR tubes that are inserted into a microwave resonator (Bruker ER4104OR, 9.43 GHz) with optical access and a helium cryostat insert (Oxford ESR900). A cw excitation source (laser or LED) on one side opening of the resonator illuminates the sample. A photodetector on the opposite opening collects the continuous PL emission with a suitable long pass filter to block residual excitation light.

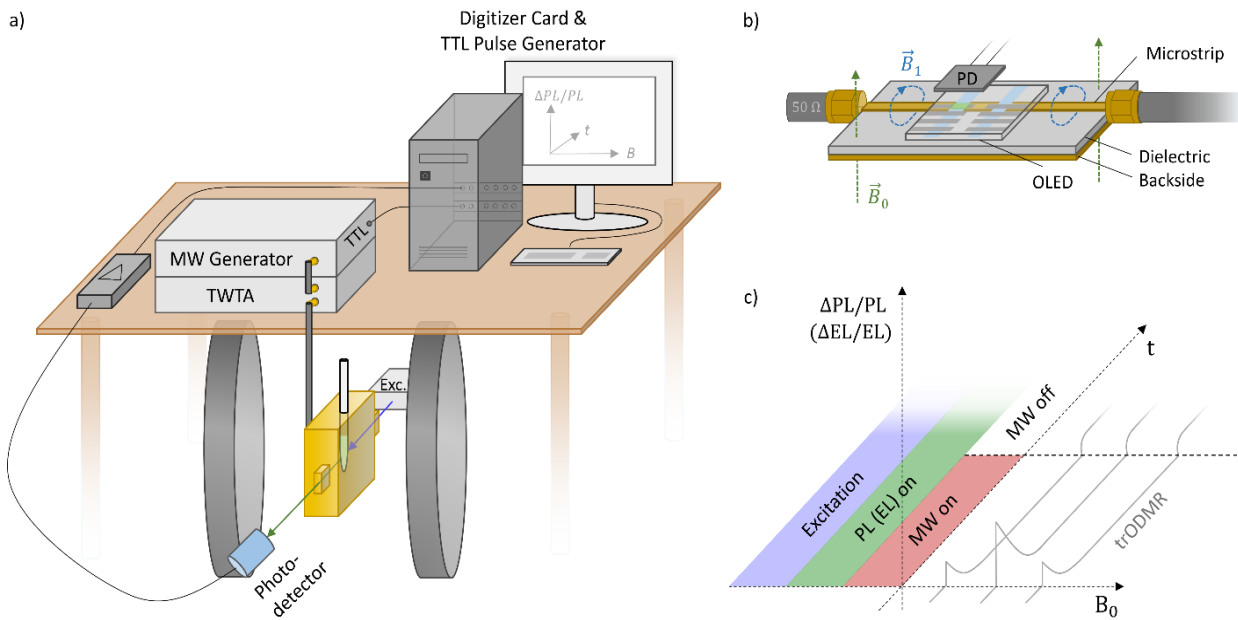

**Figure 1.** Experimental details of transient ODMR. (a) Setup for trPLDMR using a microwave resonator (gold) with openings for optical excitation and detection. PL is first amplified and then recorded by a digitizer card. A TTL pulse generator triggers the MW generator whose output is amplified by a traveling wave tube amplifier (TWTA) and guided into the resonator. (b) Microwave transmission line for trELDMR of electrically driven devices (here OLED) with photodetector (PD) for light detection. The oscillating $B_1$ MW-field crosses the device placed on top. (c) Pulse and detection sequence of trODMR. Continuous optical or electrical excitation (blue) results in continuous PL or EL (green), respectively. ODMR transients of relative change in PL or EL (grey) are recorded while microwave pulses are applied (MW on, red). The external magnetic field $B_0$ is swept to record a two-dimensional data set.

**Figure 1b** shows the microwave transmission line used for trELDMR. It consists of a metallic microstrip on a dielectric substrate with a conducting backside.[32] Similar to a coaxial cable, the electromagnetic waves propagate in the dielectric with the oscillating magnetic field component $B_1$ crossing the EL-emitting devices that is placed directly on top. This construction gives the device a high architectural flexibility and is also not limited to a unique microwave resonance frequency such as with typical EPR resonators. A source-measuring unit (Keithley 236) operates the device with a constant current to ensure continuous EL emission that is detected by a photodetector placed on top. A cryostat (Oxford 935) offers temperature control and ensures an oxygen-free atmosphere.

**Figure 1c** shows the corresponding pulse and detection sequence. Continuous optical (or electrical) excitation produces continuous PL (or EL). At a set magnetic field, microwave pulses are applied for a defined duration (MW on/off) resulting in a spin-dependent change of luminescence (ΔPL/PL, ΔEL/EL). In contrast to pulsed ODMR or EPR with the intention of coherent spin manipulation, the microwave pulse duration is longer to allow resonant effects to reach equilibrium intensities. After the microwave pulse, the system relaxes back to its previous equilibrium state of ΔPL/PL = 0 (ΔEL/EL = 0). Stepwise sweeping the external magnetic field $B_0$ between ODMR transients results in two-dimensional data sets providing time-dependent ODMR spectra. Due to direct recording of the optical signal, the spectra obtain the correct signs and amplitudes. These information are valuable, i.e. when identifying exciton loss processes or comparing spin polarization.[33] This is in stark contrast to cw techniques that usually employ lock-in detection with microwave on/off or magnetic field modulation, which results in time-averaged specta with ambigous signal signs and amplitudes depending on modulation frequency and lock-in phase.[33] The possible time-resolution of the implemented technique is limited by either the optical detection (e.g. the RC constant of the large-area (1 cm$^2$) photodetector), the bandwidth of the current-voltage amplifier or the quality factor of the microwave resonator, allowing time resolutions down to tens of nanoseconds. The microwave magnetic field strength $B_1$ is determined by the conversion factors of the used resonator ($C = 0.1$ mT W$^{-1/2}$) and transmission line ($C \approx 0.01$ mT W$^{-1/2}$).

### III. Results and discussion

**Case study: Optically- and electrically-induced triplet states**

In the following, we apply trODMR to two intensively studied organic semiconductor donor:acceptor systems whose operating principles strongly depend on excited triplet states: First, OPV systems based on state-of-the-art non-fullerene acceptors (NFA) and, second, the newest generation of OLEDs based on thermally activated delayed fluorescence (TADF). For both systems, ODMR spectra can be used as "fingerprint" for the triplet states participating in the operation mechanisms.[18]

**1. Organic Photovoltaics**

Triplet exciton formation is problematic for OPV as triplet states are notoriously difficult-to-detect exciton loss mechanisms or trap states.[34, 35] Furthermore they can result in singlet oxygen formation and are in general involved in material degradation.[18, 36, 37] cwODMR has already been shown to be well suited to study excited triplet states in such systems.[18, 35] For trODMR demonstration purposes, we choose a state-of-the-art donor:acceptor system containing the donor polymer PM6 (poly[(2,6-(4,8-bis(5-(2-ethylhexyl-3-fluoro)thiophen-2-yl)-benzo[1,2-b:4,5-b']dithiophene))-alt-(5,5-(1',3'-di-2-thienyl-5',7'-bis(2-ethylhexyl)benzo[1',2'-c:4',5'-c']dithiophene-4,8-dione)]) and the acceptor ITIC (3,9-bis(2-methylene-(3-(1,1-dicyanomethylene)-indanone))-5,5,11,11-tetrakis(4-hexylphenyl)-dithieno[2,3-d:2',3'-d']-s-indaceno[1,2 b:5,6-b']dithiophene).

**Figure 2a** shows a two-dimensional (time vs. magnetic field) trPLDMR plot of a PM6:ITIC thin film at $T$ = 10 K and excitation wavelength of 473 nm. Microwave pulses are applied from $t$ = 0 – 10 ms at a frequency of 9.43 GHz. The PLDMR transient on the right ($B$ = 336.5 mT) clearly reveals the luminescence response (ΔPL/PL) to magnetic resonance conditions. A spectral slice is extracted at the end of the MW pulse at $\Delta t$ = 9 – 10 ms, resulting in a so-called full field (FF) spectrum. It reveals a broad, wing-like signal of almost 100 mT width, corresponding to closed-by spins with a zero-field splitting of $D$ = 1300 MHz.[38] We can assign this signal to triplet excitons localized on ITIC acceptor molecules ($T_A$) as discussed in the following. First, the same zero-field splitting value is extracted for trPLDMR measurements of pristine ITIC (Figure S1). Second, the so-called half field (HF) signal (at half the magnetic field of the FF) of PM6:ITIC and pristine ITIC (Figure S2) is identical in shape and position. Since the intensity of the HF signal is strongly dependent on the inter-spin distance $r$ of the two charges bound in the triplet state (intensity scales $\sim r^{-6}$)[39], it is evident that the HF signal originates from the broad feature in the FF spectrum. The HF and the broad FF signal in PM6:ITIC can thus be assigned to triplet excitons on ITIC molecules.

The narrow (~ few mT) central peak originates from charge transfer (CT) states and non-geminate recombination of distant polaron pairs (PP), considered as spin-correlated radical pairs with negligible zero-field splitting $D$.[18, 40] The PLDMR transient on the right at $B$ = 336.5 mT shows the temporal response of the narrow peak during the MW pulse. The most recent theory is spin mixing of the singlet and triplet CT states ($^1$CT and $^3$CT) via hyperfine interaction, since the delocalization makes the spin orbit coupling interaction negligible.[41] In the case of an externally applied magnetic field, Zeeman splitting only allows spin mixing between the triplet $^3$CT$_0$ ($m_s$ = 0) and the singlet $^1$CT.[41-43] Application of microwave irradiation changes the current spin polarization of the triplet states by inducing transitions between $^3$CT$_0$, $^3$CT$_{-1}$ and $^3$CT$_{+1}$ states. The positive sign (ΔPL/PL > 0) suggests repopulation of the light-emitting channel in resonant conditions. A new equilibrium intensity is then reached within about 10 ms. After switching off the MW, the PL intensity relaxes back with a similar time constant. High modulation frequencies ($f_{mod}$ > 100 Hz) in cwODMR would accordingly misrepresent the measurement result.

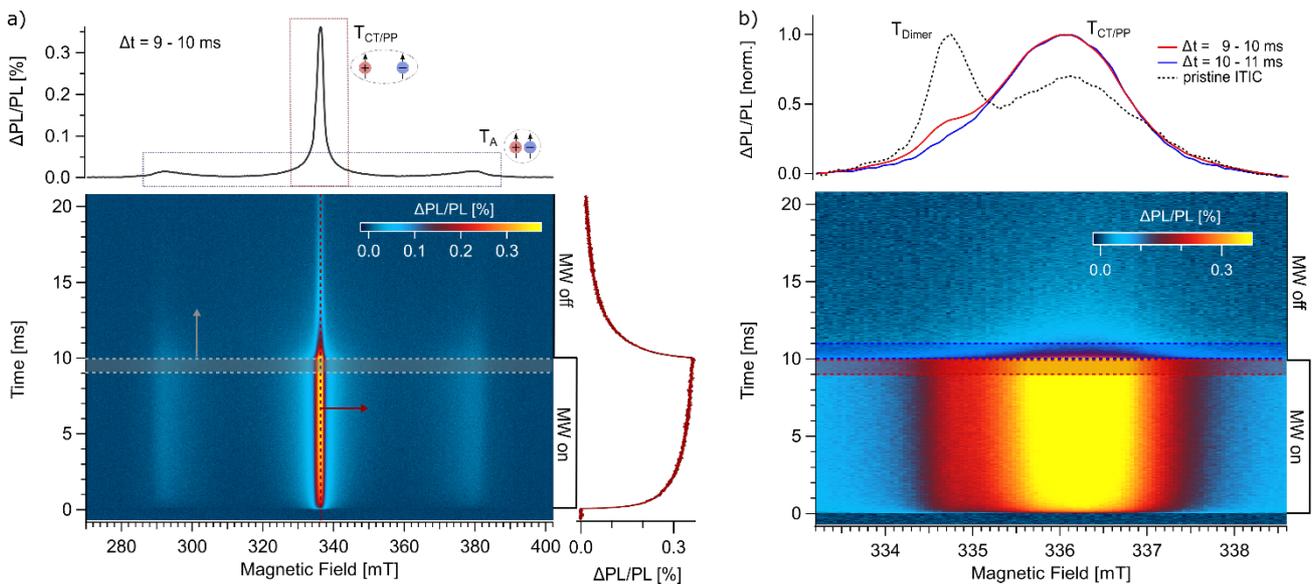

**Figure 2.** trPLDMR of the OPV donor:acceptor system PM6:ITIC at $T$ = 10 K. (a) Two-dimensional full field (FF) plot. A horizontal slice from $\Delta t$ = 9 – 10 ms (grey spectrum, top) shows a PLDMR spectrum with contributions of a broad wing-like feature due to ITIC triplets ($T_A$) and a central peak due to charge transfer states or polaron pairs ($T_{CT/PP}$). The PLDMR transient (red trace, right) for $B$ = 336.5 mT is recorded when the resonant MW pulse is switched on and off. (b) Central $T_{CT/PP}$ peak with higher spectral resolution. An additional shoulder at $B$ = 334.8 mT (red spectrum, top, $\Delta t$ = 9 – 10 ms) is revealed that is separable from the CT/PP peak, since this signal vanishes quickly after the MW pulse (blue spectrum, top, $\Delta t$ = 10 – 11 ms). This contribution ($T_{Dimer}$) is also observed for pristine ITIC (black, dashed spectrum, top, $\Delta t$ = 9 – 10 ms from Figure S1).

Time-dependent ODMR spectra are also useful for separating superimposed signals. **Figure 2b** shows the narrow CT/PP peak with higher spectral resolution. Additional to the main central peak ($T_{CT/PP}$), there is a discernible shoulder at around $B$ = 334.8 mT (red normalized spectrum, averaged over $\Delta t$ = 9 – 10 ms). This signal is separable from the CT/PP peak itself, since it vanishes quickly after microwave switch off (blue normalized spectrum, averaged over $\Delta t$ = 10 – 11 ms). This analysis is corroborated by detecting the same spectral contributions in pristine ITIC (grey, dashed normalized spectrum, averaged over $\Delta t$ = 9 – 10 ms from Figure S1). Previously, the shoulder could be tentatively assigned to an additional triplet exciton delocalized over ITIC dimers ($T_{Dimer}$).[44] It is not as pronounced in the blend since additional donor:acceptor CT states contribute to the CT/PP peak at 336.5 mT and blend mixing decreases the impact of ITIC dimers on the overall signal. Nevertheless, this additional contribution is conclusively detectable and separable with trODMR.

**2. Organic light emitting diodes**

In the trODMR spectra of the above-discussed OPV systems, the CT/PP peak is quite narrow ($D \to 0$) and easily distinguishable from the broad ($D$ = 1300 MHz) molecular triplet feature. In other systems, the involved triplet species do not exhibit such a significant difference in zero-field splitting $D$, making their separation more challenging due to significant overlap. trODMR enables to separate spectral and temporal behavior of these different spin species, as demonstrated in the following on a TADF OLED system. In addition to trPLDMR of optically excited films, we use trELDMR on electrically driven devices to reveal crucial differences in the spin states involved and thus the importance of spin-sensitive spectroscopy for both PL and EL simultaneously.

The newest generation of OLEDs is based on harvesting non-emissive triplet excitons using thermal energy to overcome the small energy gap to the emissive singlet state.[45, 46] Subsequently, the thermally activated triplet excitons contribute with delayed fluorescence to luminescence, enhancing the quantum efficiency significantly. For the demonstration of trODMR, we choose the TADF donor:acceptor combination m-MTDATA:3TPYMB with 4,4′,4′′-tris[phenyl(m-tolyl)amino]triphenylamine as donor and tris(2,4,6-trimethyl-3-(pyridin-3yl)phenyl)borane as acceptor. This model system attracts a lot of attention in literature based on a great variety of measurement techniques, including magnetic field dependent studies[42, 47, 48]. **Figure 3a** shows the two-dimensional trPLDMR plot of the donor:acceptor blend film with 365 nm optical LED excitation at $T$ = 50 K. Microwave pulses are applied from $t$ = 0 – 1.5 ms, which is sufficient to reach equilibrium intensities. An in-resonance transient is shown on the right (red, $B$ = 336.5 mT). It reveals a fast rising (~ 0.1 ms), positive peak ($\Delta$PL/PL > 0), followed by an exponential decay to reach equilibrium with negative intensity ($\Delta$PL/PL < 0). This behavior already indicates superimposed signals with opposite sign of different spin states involved in PL. The spectrum on the top (grey) shows a slice of the 2D data set, averaged between $\Delta t$ = 0.25 – 0.35 ms. This spectrum emphasizes the observation of two superimposed signals by revealing a broad positive and a narrow negative signal with the same center magnetic field.

The width of the spectrum contains information about the delocalization of the spin species ($D \sim r^{-3}$).[18, 38] Consequently, the broad signal represents a more localized while the narrow signal represents a more delocalized triplet state. The 2D data set shows the temporal development of these two signals: the broad signal is predominantly present at the beginning of the microwave pulse while the narrow, negative signal dominates at later times. After switching off the microwave irradiation at 1.5 ms, both signals relax with different time constants and opposite sign back to their previous stationary state.

Hereby, trODMR shows its strength by giving access to a range of spectra for different points in time. An appropriate two-dimensional analysis is used to deconvolute the superimposed spectra and determine the individual transient intensities, i.e., signal contributions. Therefore, we make the following assumptions: the signal origins, e.g. the triplet states (≙ line widths) remain identical over time, only the amplitudes change. We can therefore apply a global fit with two Gaussian line shapes (which represent the envelopes of inhomogeneously broadened spectra) to these time-dependent spectra. The top graph of Figure 3a shows the superimposed spectra, consisting of a narrow spectrum with a line width (FWHM) of $\Delta B_1$ = 2.7 mT (blue dotted line) and a broader contribution with $\Delta B_2$ = 6.6 mT (red dotted line). The deconvoluted temporal amplitudes of these two triplet states are shown on the right: The narrow signal rises slowly to a negative value of about 0.1% (blue dotted line), while the broad signal presents a steep rising peak followed by a slow decay (red dotted line). The sum of these amplitudes (black line) fits the in-resonant transient. The same analysis for measurements up to 200 K is given in the Supporting Information (Figure S3) revealing that both triplet states are involved over the full temperature range.

PL-based measurement methods are well suited to obtain valuable information about the material systems themselves. However, OLED materials are intended to be used in electrically driven devices where electric fields and injected charge carriers may result in very different conditions than optical excitation. Therefore, for the study of operational devices, we demonstrate trELDMR on an OLED with an emissive layer of the same material system. The device structure is ITO/PEDOT:PSS/m-MTDATA (30 nm)/m-MTDATA:3TPYMB (70 nm, 1:1)/3TPYMB (30 nm)/Ca (5 nm)/Al (120 nm). Details about these devices are published elsewhere.[20]

**Figure 3b** presents the two-dimensional trELDMR results at $T$ = 200K. The MW pulse is applied from $t$ = 0 – 0.5 ms to reach equilibrium signal intensities. The use of a microwave transmission line makes the choice of the resonant frequency variable. Therefore, we used 8 GHz due to good signal intensities. The extracted in-resonance transient on the right (red) reveals just one negative signal with exponentially decaying intensity upon microwave pulse start/end. The spectrum on the top (grey) is averaged between $\Delta t$ = 0.4 – 0.5 ms and displays a Gaussian line shape with negative sign. For direct comparison with trPLDMR measurements from Figure 3a, we again performed a two-dimensional analysis with a single Gaussian fit function. The resulting line width of $\Delta B_1$ = 3.0 mT (top, blue dotted line) is almost identical to the narrow line width from the trPLDMR analysis. Furthermore, the temporal amplitude of this signal (right, blue dotted) matches with the measured in-resonance transient which proofs that just one triplet state is probed.

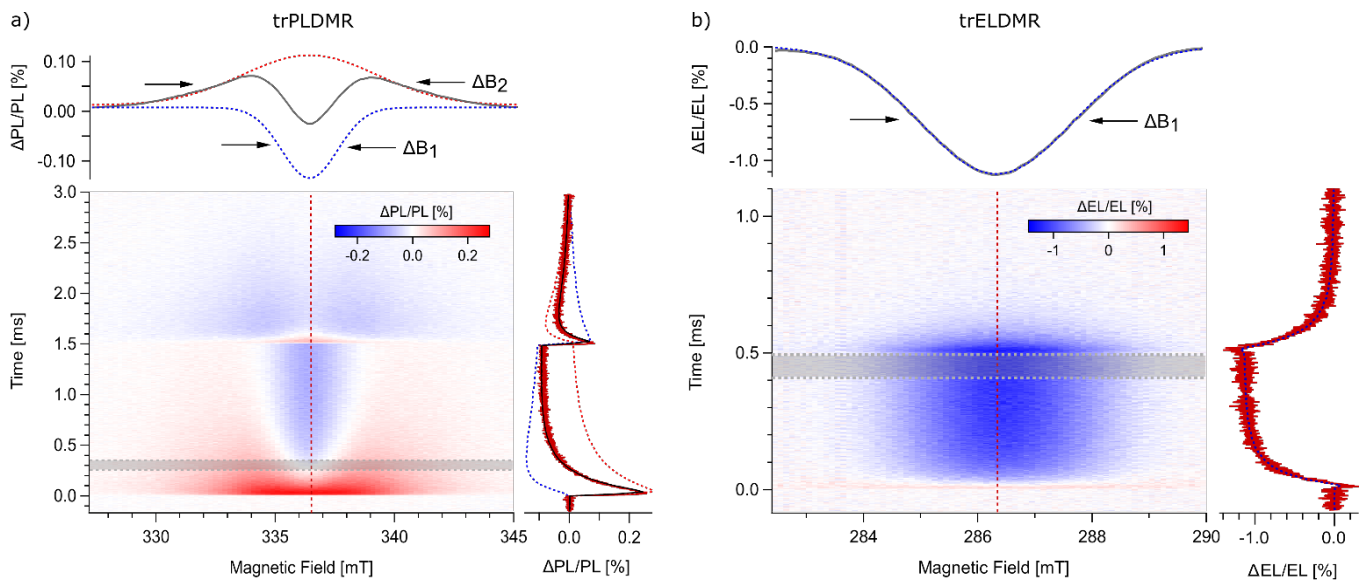

**Figure 3.** Transient ODMR of the OLED donor:acceptor system m-MTDATA:3TPYMB. (a) Two-dimensional trPLDMR plot of an optically excited thin film sample at $T$ = 50 K. A PLDMR spectrum averaged from $\Delta t$ = 0.25 – 0.35 ms (grey trace, top) and a PLDMR transient at $B$ = 336.5 mT (red trace, right) reveal two superimposed signals. By applying a global fit of two Gaussian line widths, a narrow ($\Delta B_1$, blue dotted) and broad ($\Delta B_2$, red dotted) spectrum with their temporal contributions (dotted lines, right) can be distinguished. (b) Two-dimensional trELDMR plot of an electrically driven OLED device at $T$ = 200 K. An ELDMR spectrum averaged from $\Delta t$ = 0.4 – 0.5 ms (grey trace, top) and an ELDMR transient at $B$ = 286.3 mT (red trace, right) reveal just one Gaussian signal ($\Delta B_1$), supported by the results of the global fit (blue dotted lines, top/right).

This delocalized triplet state (narrow signal) is thus involved in both, electrically driven OLED systems and optically excited thin films. In contrast to that, trPLDMR of optically excited films reveals a second, broader spectral component, which also exists up to 200 K as shown in Figure S3. The larger width is caused by stronger spin-spin interaction of a more localized triplet state, only present in PL emitting thin films. While electrical excitation results in delocalized intermolecular CT states, optical excitation most probably involves additional more localized CT states, e.g., distributed over nearest neighbor molecules. Therefore, experiments based on optical excitation must be treated with special caution: The emitting final state may be identical, but there may be different intermediate excited states and recombination pathways that are absent for electrical excitation. However, they can be resolved by trPLDMR and, in comparison with trELDMR, be distinguished from spin states accessible in current-driven devices.

## IV. Conclusions

We demonstrate the method of transient optically detected magnetic resonance (trODMR), which is a spectrally and time resolved technique that directly probes the spin-carrying excited states participating in photo- or electroluminescence. This technique allows deconvolution and identification of superimposed spectral components occurring on different time scales with the superior sensitivity of optical detection. The method can be applied to all materials with spin-dependent luminescence, including systems with fast spin-spin relaxation times and especially operating opto-electronic devices. In two case studies, we apply this technique to molecular donor:acceptor systems widely used in organic photovoltaics based on non-fullerene acceptors (NFA) and OLEDs based on thermally activated delayed fluorescence (TADF). A global fit analysis of the time-dependent ODMR spectra provides the exact spectral contributions with absolute sign and amplitude and thus reveals the individual spin-dependent radiative recombination paths. In particular, we find that different intermediate triplet states are involved in optically excited films and electrically driven devices, which has far-reaching implications for materials research, since this decisive difference is often undetectable with spin-insensitive methods.

**Conflicts of interest**

There are no conflicts to declare.

**Acknowledgements**

This work was supported by the Deutsche Forschungsgemeinschaft (DFG, German Research Foundation) within the Research Training School "Molecular biradicals: Structure, properties and reactivity" (GRK2112).

SUPPORTING INFORMATION

for

# Detecting triplet states in opto-electronic and photovoltaic materials and devices by transient optically detected magnetic resonance

Jeannine Grüne, Vladimir Dyakonov and Andreas Sperlich*

Experimental Physics 6, Julius Maximilian University of Würzburg,
Am Hubland, 97074 Würzburg, Germany

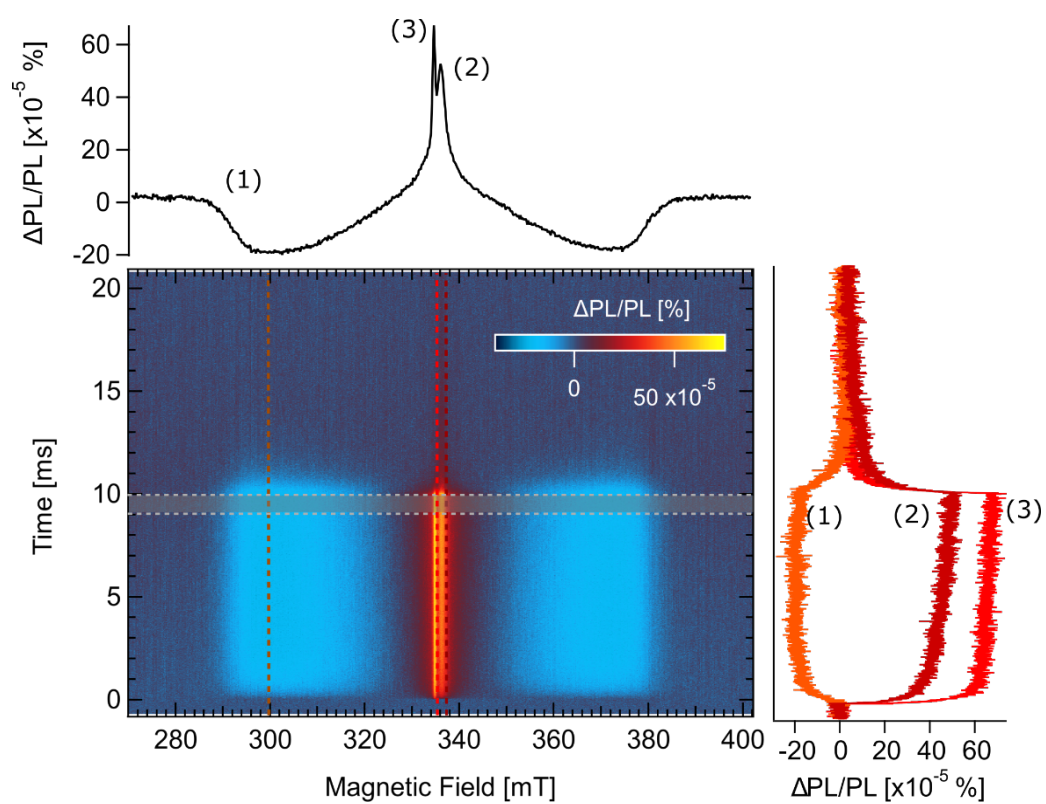

**Figure S1.** Transient PLDMR of a pristine ITIC film with exemplary transients (right) and spectrum (top), averaged over 9 – 10 ms. There are three superimposed spectral contributions: (1) A wide "wing-like" pattern (~100 mT wide, $D$ = 1300 MHz) that can be assigned to molecular ITIC triplet excitons. (2) A central CT/PP peak at $B$ = 336.5 mT with (3) an additional narrow spike at 334.8 mT. Measured with 473 nm laser excitation at $T$ = 10 K and MW pulse length of 10 ms.

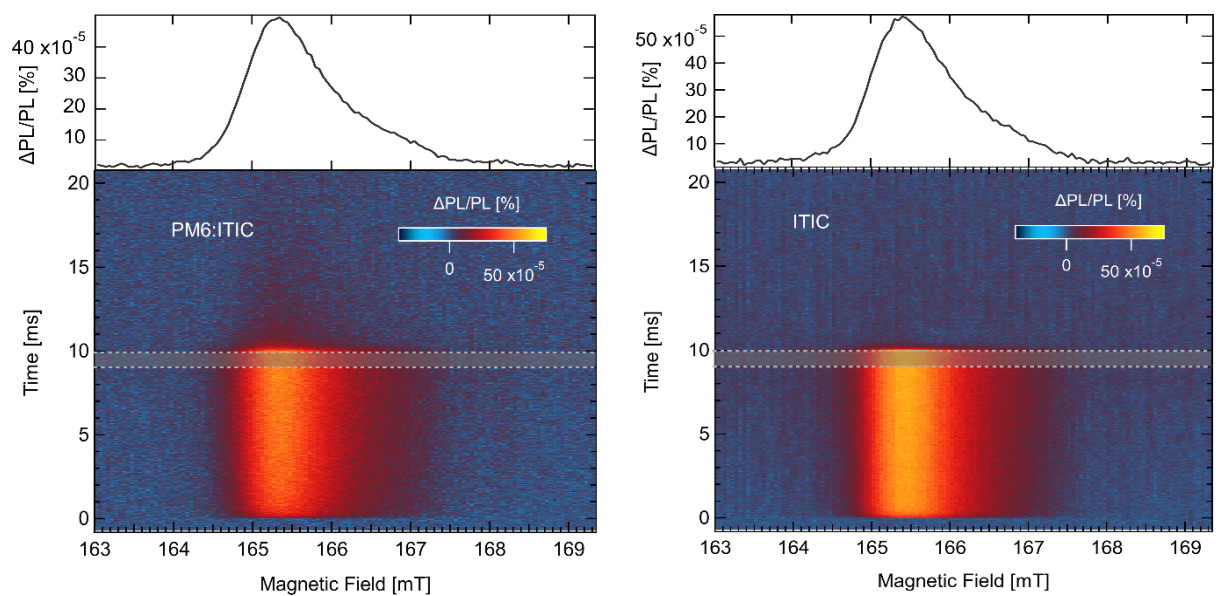

**Figure S2.** Transient PLDMR of the halffield (HF) signals of OPV blend PM6:ITIC blend (left) and pristine ITIC (right). Averaged spectra from 9 – 10 ms during microwave pulse are shown on top. Both signals have the same spectral position (g-factor), shape and comparable intensities. Measured with 473 nm laser excitation at $T$ = 10 K and MW pulse length of 10 ms.

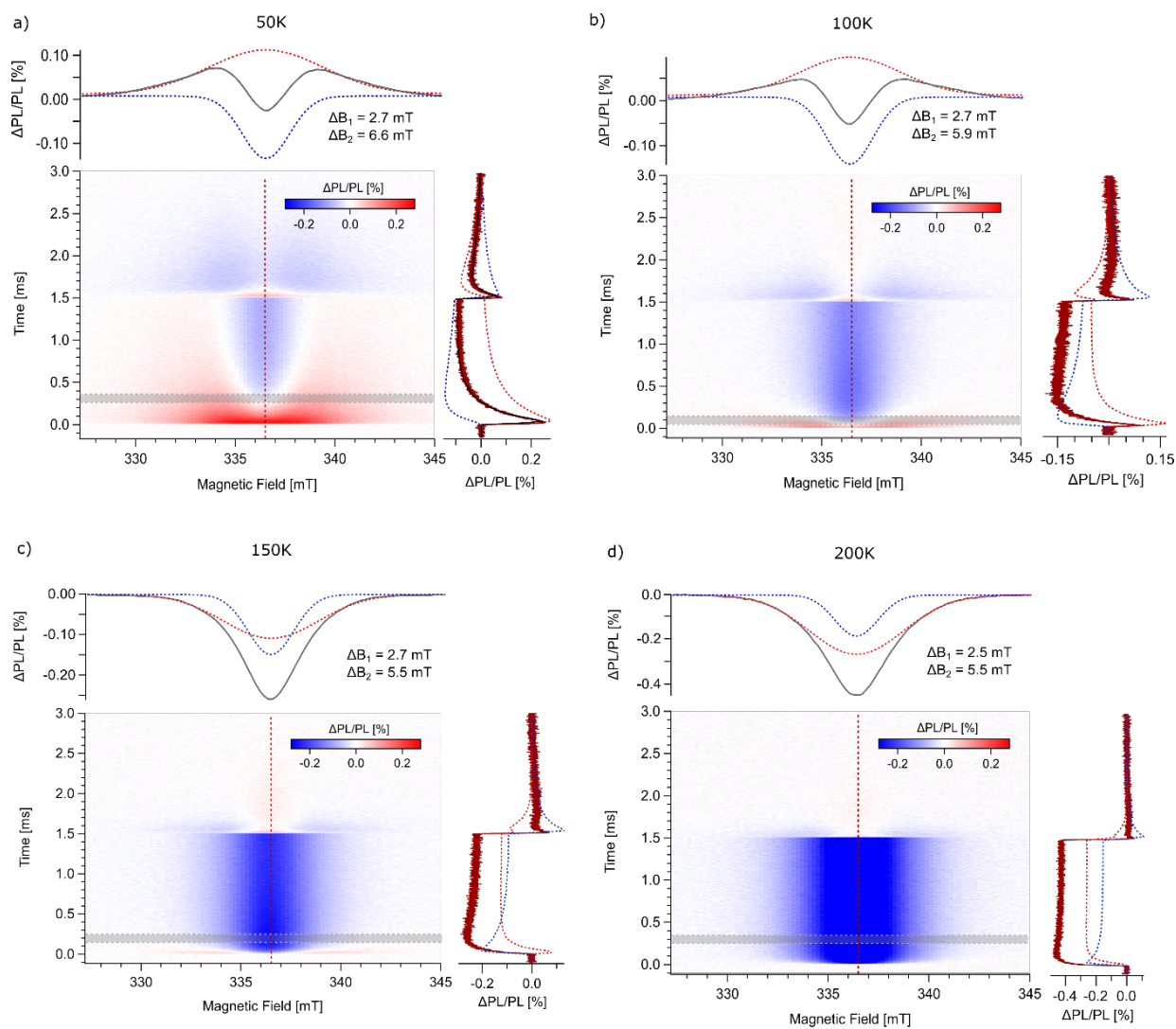

**Figure S3.** Transient PLDMR of OLED m-MTDATA:3TPYMB blend for 50 K, 100 K, 150 K and 200 K. During the 1.5 ms long microwave pulses, first, strong perturbations with changing signs are observed, before the intensities reach a new equilibrium under these conditions. Opposite trends are observed upon switching off the MW pulse. At all temperatures, signals of two triplet states and their time-dependent intensity traces can be extracted by a global fit with two Gaussian FWHM linewidths ($\Delta B_1$, $\Delta B_2$). The contribution of the narrower triplet signal $\Delta B_1$ (blue) is always negative and switches sign afterwards. The contribution of the broader triplet signal $\Delta B_2$ (red) is positive during the MW pulse at low temperatures and becomes more negative with increasing temperature or time. Measured with 365 nm LED excitation.